%% file: main.tex
\documentclass[pdflatex,sn-nature]{sn-jnl}

\usepackage{graphicx}%
\usepackage{multirow}%
\usepackage{amsmath,amssymb,amsfonts}%
\usepackage{amsthm}%
\usepackage{mathrsfs}%
\usepackage[title]{appendix}%
\usepackage{xcolor}%
\usepackage{textcomp}%
\usepackage{manyfoot}%
\usepackage{booktabs}%
\usepackage{algorithm}%
\usepackage{algorithmicx}%
\usepackage{algpseudocode}%
\usepackage{listings}%

\usepackage[braket]{qcircuit}

\usepackage{tikz}
\usepackage{amsmath}
\usetikzlibrary{arrows.meta}

\theoremstyle{thmstyleone}%
\theoremstyle{thmstyletwo}%

\theoremstyle{thmstylethree}%

\begin{document}

\title[Noise-Resilient Spatial Search with Lackadaisical Quantum Walks]{Noise-Resilient Spatial Search\par with Lackadaisical Quantum Walks}


\author[1]{\fnm{Gabriel} \sur{Mauricio Oswald Vieira}}\email{gvieira@cos.ufrj.br}

\author[1]{\fnm{Nelson} \sur{Maculan}}\email{maculan@cos.ufrj.br}

\author*[1,2]{\fnm{Franklin} \spfx{de} \sur{Lima Marquezino}}\email{franklin@cos.ufrj.br}

\affil*[1]{\orgdiv{Systems Engineering and Computer Science Program (COPPE)}, \orgname{Federal University of Rio de Janeiro}, \orgaddress{\street{Avenida Horácio Macedo 2030, CT-H319}, \city{Rio de Janeiro}, \postcode{21941-914}, \state{RJ}, \country{Brazil}}}

\affil[2]{\orgdiv{Duque de Caxias Campus}, \orgname{Federal University of Rio de Janeiro}, \orgaddress{\street{Rodovia Washington Luiz, 19593}, \city{Duque de Caxias}, \postcode{25240-005}, \state{RJ}, \country{Brazil}}}


\abstract{Quantum walks are a powerful framework for the development of quantum algorithms, with lackadaisical quantum walks (LQWs) standing out as an efficient model for spatial search. In this work, we investigate how broken-link decoherence affects the performance of LQW-based search on a two-dimensional toroidal grid. We show through numerical simulations that, while decoherence drives the loopless walk toward a uniform distribution and eliminates its search capability, the inclusion of self-loops significantly mitigates this effect. In particular, even under noise, the marked vertex remains identifiable with probability well above uniform, demonstrating that self-loops enhance the robustness of LQWs in realistic scenarios. These findings extend the known advantages of LQWs from the noiseless setting to noisy environments, consolidating self-loops as a valuable resource for designing resilient quantum search algorithms.}




\maketitle

\section{Introduction}


Quantum walks have been investigated for more than three decades~\cite{aharonov_quantum_1993}, emerging as one of the central tools for the design and analysis of quantum algorithms. They provide a natural quantum counterpart to classical random walks, and this connection has inspired a wide range of algorithmic applications. Among their most notable contributions are quantum search~\cite{ambainis_coins_2005, portugal_quantum_2018} and optimization algorithms, where quantum walks enable significant speedups over classical strategies. In fact, quantum walks can even achieve exponential speedup in specific oracle problems, such as the welded trees traversal~\cite{childs_exponential_2003}, highlighting their algorithmic power beyond quadratic improvements. Beyond these concrete applications, quantum walks are also known to be universal for quantum computation~\cite{childs_universal_2009, lovett_universal_2010}, consolidating their role as a unifying framework within quantum information science.

In the specific setting of spatial search on a two-dimensional grid, a standard reference point is the quantum walk algorithm introduced by Ambainis, Kempe, and Rivosh (AKR)~\cite{ambainis_coins_2005}. The AKR algorithm employs a discrete-time coined quantum walk with the Grover coin applied at all unmarked vertices, a phase-flip coin $-I$ at the marked vertex, and the flip-flop shift, achieving a runtime of $O(\sqrt{N\log N})$ with success probability $O(1/\log N)$. An important refinement was later introduced by Tulsi, who proposed an ancilla-controlled quantum walk that amplifies the success probability to a constant while preserving the $O(\sqrt{N\log N})$ runtime~\cite{tulsi_faster_2008}.

The motivation for studying lazy walks, where the walker has some probability of staying put, comes from the fact that it is a useful tool in classical computing~\cite{wong_grover_2015}. 
Lazy walks can be constructed in different ways. Childs~\cite{childs_relationship_2010} proposed one such formulation in the context of relating continuous- and discrete-time quantum walks, while Wong~\cite{wong_grover_2015} later introduced the lackadaisical quantum walk by adding weighted self-loops to each vertex of the original graph.
 In this work, we will consider lackadaisical discrete-time quantum walks. 
 The lackadaisical approach for quantum lazy walks is illustrated in Fig.~\ref{fig:lackdaisical}, where each vertex of the two-dimensional grid is equipped with a self-loop of weight $\ell$, allowing the walker not only to move to neighboring sites but also to remain at the same vertex; the white circle indicates the marked vertex used in the subsequent search analysis.

\begin{figure}[hbt]
\begin{center}

\input{lackadaisical.tikz}

\end{center}
\caption{\label{fig:lackdaisical}A two-dimensional grid with $4\times 4$ vertices, self-loops of weight $\ell$ on each vertex, and periodic boundary. The marked vertex is indicated by a white circle.}
\end{figure}

An important contribution to the study of quantum walks with self-loops was presented by Høyer and Yu~\cite{hoyer_analysis_2020}, who provided a rigorous analytical treatment of  lackadaisical quantum walks. They showed that, by adding a self-loop of appropriate weight to each vertex of a locally arc-transitive regular graph, it is possible to find a single marked vertex with constant probability in time $O(\sqrt{\mathsf{HT}})$, where $\mathsf{HT}$ is the classical hitting time. Their analysis establishes a formal equivalence between LQWs and Szegedy's interpolated quantum walks, enabling the use of well-established analytical tools. This result confirms and generalizes several previous conjectures based solely on numerical evidence, such as in the cases of the 2D torus, cycles, and hypercubes, thus consolidating LQWs as an efficient approach for search algorithms on structured graphs.

Rhodes and Wong provided numerical evidence that lackadaisical quantum walk search with a single marked vertex achieves optimal performance on a broad class of vertex-transitive graphs, with the optimal self-loop weight scaling as the degree of the underlying graph divided by the number of vertices~\cite{rhodes_search_2020}.
More recently, lackadaisical quantum walks have been shown to achieve asymptotically unit success probability for spatial search on other structured graph families, such as strongly regular graphs, with rigorous analytical guarantees~\cite{peng_search_2024}. Related extensions to other highly symmetric graph families, as well as to scenarios involving multiple marked vertices, have also been explored in recent work~\cite{giri_multi-target_2025}. In addition, numerical studies have investigated the performance of lackadaisical quantum walk search in the presence of multiple or clustered marked vertices, revealing that appropriately tuned self-loop weights can yield near-unit success probability in these settings~\cite{giri_lackadaisical_2019, saha_faster_2022}.


Studying decoherence in quantum walks is crucial for moving from idealized models to practical algorithms. Since real quantum devices are unavoidably noisy, identifying mechanisms that preserve performance, such as the inclusion of self-loops, helps in designing quantum-walk-based algorithms that remain effective under realistic conditions.


In this work, we present numerical evidence that lackadaisical quantum walks are more robust for spatial search in the presence of structural noise modeled by dynamic broken links. By comparing loopless and lackadaisical walks on a two-dimensional toroidal grid, we show that self-loops mitigate the degradation caused by decoherence, allowing the marked vertex to remain identifiable even when noise drives the loopless walk toward uniformity. Our analysis highlights success probability and long-term average distributions as key indicators of this robustness. These results not only extend previous studies on lackadaisical walks in the noiseless case but also provide insights into how to design quantum-walk-based algorithms that are better suited for near-term quantum devices.

This work is organized as follows. In Sec.~\ref{sec:lackadaisical}, we briefly review lackadaisical quantum walks and introduce notation. In Sec.~\ref{sec:decoherence}, we discuss our decoherence model. In Sec.~\ref{sec:numerical}, we present our main results. In Sec.~\ref{sec:conclusions}, we present our conclusions and final discussions.

\section{Lackadaisical quantum walks}
\label{sec:lackadaisical}


Discrete-time quantum walks (DTQWs) were first defined by Aharonov et al.~\cite{aharonov_quantum_1993}, and they provide a versatile framework for exploring quantum dynamics on graphs. Unlike their classical counterparts, where a walker probabilistically moves along vertices of the graph, a quantum walker evolves unitarily, maintaining superposition and coherence across different paths. To model this, an auxiliary Hilbert space --- the coin space --- is introduced, encoding the possible directions in which the walker may move. Therefore, the state of the walker will be $\ket{\psi} \in \mathcal{H}_C \otimes \mathcal{H}_P,$ where $\mathcal{H}_C$ is the coin space, and $\mathcal{H}_P$ is the position space.

The evolution of the walk is then defined by two unitary operators. The coin operator $C$ acts locally on the coin space, creating a superposition of directions, while the shift operator $S$ conditionally moves the walker to adjacent vertices based on the coin state. A single step of the walk is described by the unitary operator
\begin{equation}
U = S \,(C \otimes I),
\end{equation}
where $I$ acts on the position space. A widely used choice for $C$ is the Grover diffusion operator~\cite{ambainis_coins_2005},
\begin{equation}
C = 2 \ket{s_c}\bra{s_c} - I_d,
\label{eq:grover_coin}
\end{equation}
where $d$ is the degree of the graph, $\ket{s_c}$ is the uniform superposition of coin states, and $I_d$ is the identity. This idea can be easily generalized to non-regular graphs by applying a different coin for each vertex.


Repeated applications of $U$ generate interference patterns that differ markedly from classical diffusion and form the basis of algorithmic applications such as spatial search. In the context of searching algorithms, the shift operator is often taken to be the flip-flop shift, which not only moves the walker from vertex $u$ to a neighboring vertex $v$, but also flips the coin state to point back from $v$ toward $u$~\cite{ambainis_coins_2005}. 
In order to turn the walk into a spatial search algorithm, an additional operator is required to mark the target vertices. This is achieved by applying the reflection
\begin{equation}
R = I - 2\sum_{m \in M}\ket{m}\bra{m},
\end{equation}
which flips the phase of the marked states, where $M$ denotes the set of marked vertices. For simplicity, we restrict our analysis to the case $|M|=1$, with the marked vertex chosen at the center of the grid without loss of generality. The search algorithm is then implemented by repeatedly applying the combined evolution operator $U^\prime = RU$, which drives amplitude toward the marked vertex through quantum interference.


In order to define a lackadaisical quantum walk, the coin operator is modified by introducing an additional coin state that corresponds to a self-loop at each vertex~\cite{wong_grover_2015}. Concretely, this is done by replacing the uniform coin state in Eq.~\ref{eq:grover_coin} with
\begin{equation}
\ket{s_c} \;=\; \frac{1}{\sqrt{d+\ell}} \left( \sum_{i=0}^{d-1}\ket{i} + \sqrt{\ell}\,\ket{d} \right),
\end{equation}
where $\ell$ is a real parameter that controls the weight of the self-loop. When $\ell=0$, this reduces to the standard Grover coin, while larger values of $\ell$ increase the amplitude associated with remaining at the current vertex. The corresponding shift operator must also be defined accordingly, so that the extra coin state $\ket{d}$ corresponds to the walker remaining at the same vertex. This construction formalizes the lackadaisical quantum walk introduced by Wong~\cite{wong_grover_2015} and later analyzed rigorously by Høyer and Yu~\cite{hoyer_analysis_2020}.

\begin{figure}[hbt]
\begin{center}

\includegraphics[width=0.9\columnwidth]{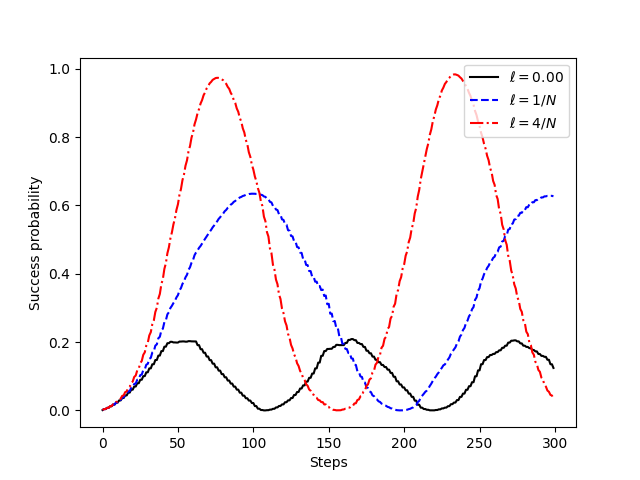}

\end{center}
\caption{\label{fig:optimal-self-loop-noiseless} Success probability as a function of time for noiseless search on the two-dimensional grid of $N=32\times 32$ vertices using a coined discrete-time quantum walk with self-loops at each vertex, for different self-loop weights $\ell$. The case $\ell=0$ corresponds to the standard Grover-coin walk, while $\ell>0$ yields the lackadaisical walk. (Color online.)}
\end{figure}

In Fig.~\ref{fig:optimal-self-loop-noiseless}, we have the evolution of the success probability for a noiseless lackadaisical quantum walk on a two-dimensional grid with $N=32\times 32$ vertices, comparing different self-loop weights $\ell$. The plot shows that, when $\ell = 4/N$, the probability of finding the marked vertex rapidly increases and saturates at a constant value close to one. For other values of $\ell$, the success probability is either significantly lower or requires a longer time to build up. This behavior is consistent with the numerical observations reported by Wong for lackadaisical quantum walk search on the two-dimensional grid~\cite{wong_faster_2018}.

\section{Percolation decoherence in discrete-time quantum walks}
\label{sec:decoherence}

The study of noise and decoherence in quantum walks is essential for bridging the gap between idealized models and practical implementations~\cite{abal_decoherence_2009}. While quantum walks can offer significant algorithmic advantages in noiseless settings, real quantum devices are inherently affected by imperfections such as gate errors, qubit relaxation, and environmental interactions. Understanding how decoherence impacts quantum-walk dynamics is therefore a crucial step toward designing algorithms that are not only theoretically efficient but also robust enough to operate on near-term quantum hardware. In this context, analyzing different noise models provides valuable insights into the resilience of quantum walks and highlights strategies --- such as the inclusion of self-loops --- that can mitigate detrimental effects and extend their applicability.

The study of quantum walks under noise has attracted considerable attention, since decoherence strongly affects their algorithmic performance. Several noise models have been considered in the literature, each capturing different physical mechanisms. A common approach is to apply quantum channels such as depolarizing or phase-damping noise to the coin or position space, which gradually suppress quantum interference and drive the walk toward classical diffusive behavior. Another important line of study involves structural noise, where the underlying graph itself is perturbed at each step. Examples include static or dynamic percolation, in which edges are randomly removed either once at the beginning or independently at every time step~\cite{abal_decoherence_2009, oliveira_decoherence_2006, santos_decoherence_2022}. In this work, we will consider dynamic percolation noise.

In Fig.~\ref{fig:broken-link-types}, we illustrate how the elimination of an edge affects the amplitude propagation in a coined discrete-time quantum walk. For simplicity, the figure omits self-loops and does not depict the flip-flop shift, but the same ideas apply when these elements are present, such as in lackadaisical quantum searches. The redirection mechanism illustrated here can be formalized within a linear-algebraic framework based on permutation operators that preserve unitarity under edge removal, as developed in~\cite{vieira_algebraic_2024}. Each circle represents a vertex, and each arrow corresponds to one component of the walker's quantum state: the joint state $\ket{v, d}$, where $v$ is the vertex and $d$ indicates the direction (or coin state) in which the walker is pointing. In part~(a), no edge has been removed, and amplitude can flow freely between adjacent vertices in both directions. The walker at vertex $v$ with coin state pointing to $w$ will hop to $w$ in the next step, and vice versa. In part~(b), edge $(v,w)$ has been removed, which prevents the walker from transitioning between those vertices. As a consequence, the amplitude that would have flowed through the removed edge must be redirected. To maintain unitarity, the amplitude is reallocated to other available coin directions at vertex $v$. This mechanism models percolation-type decoherence in the quantum walk: at each time step, the graph connectivity is dynamically modified, but the overall evolution remains unitary by adjusting the internal redistribution of amplitude~\cite{leung_coined_2010, oliveira_decoherence_2006, elster_quantum_2015, rigovacca_two-walker_2016}.

\begin{figure}[hbt]
\begin{center}

\input{broken-links.tikz}

\end{center}
\caption{\label{fig:broken-link-types} Schematic illustration of how broken links affect amplitude propagation in a coined discrete-time quantum walk. 
The elimination of an edge causes the corresponding coin-state amplitude to be redistributed among the remaining available directions to preserve unitarity. 
(a) No edge has been removed. (b) Edge $(v,w)$ has been removed.}
\end{figure}

\begin{figure}[hbt]
\begin{center}
\includegraphics[width=0.95\columnwidth]{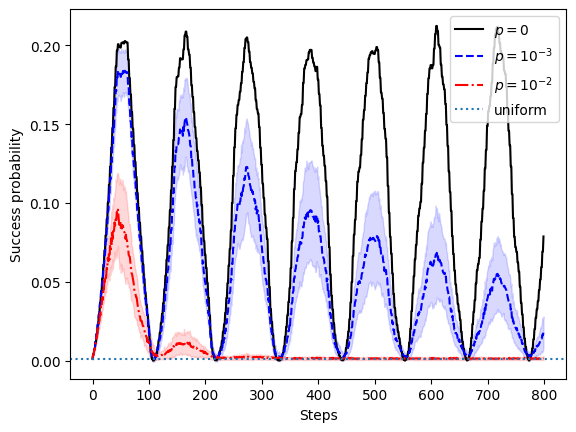}
\end{center}
\caption{\label{fig:succprob-bl}
Success probability as a function of time for the loopless walk ($\ell=0$) on a $32\times 32$ toroidal grid under broken-link decoherence, for several values of the broken-link probability $p$. Each curve represents the average over 50 independent noisy realizations, with the shaded region indicating the corresponding standard deviation. The dotted horizontal line denotes the uniform distribution baseline. (Color online.)}
\end{figure}

In Fig.~\ref{fig:succprob-bl}, we have the success probability as a function of time under broken-link decoherence, for different values of the self-loop weight $\ell$. The results indicate that self-loops play a protective role: while noise inevitably reduces the maximum success probability, an appropriate choice of $\ell$ mitigates this decay and helps maintain a clear separation from the uniform distribution baseline.

In Fig.~\ref{fig:succprob-bl}, we show the success probability as a function of time for the loopless walk $(\ell=0)$ under different broken-link probabilities 
$p$. The results indicate that increasing $p$ suppresses the peak success probability and drives the dynamics toward the uniform baseline.
For evolution times on the order of $1/p$, the quantum walk becomes effectively
fully decohered, and the search process is completely degraded, with the probability
distribution approaching uniform and the marked vertex no longer being
distinguishable~\cite{oliveira_decoherence_2006}.

\section{Numerical results}
\label{sec:numerical}

In this work, we performed numerical simulations of the lackadaisical discrete-time quantum walk on a two-dimensional toroidal grid. The walk was implemented with the Grover coin and the flip-flop shift operator~\cite{ambainis_coins_2005}, and the marked element was chosen as the central vertex of the grid, without loss of generality. We investigated the effects of varying the self-loop weight $\ell$ and the broken-link probability $p$, where decoherence was modeled as a dynamic percolation process in which edges are randomly broken and re-sampled at each step. Since the dynamics are probabilistic, all reported quantities were obtained from averages over multiple independent runs, and the standard deviation is represented in the plots as shaded regions. The grid size, number of runs, and total number of steps are specified in the captions of each figure. Our analysis focused on two main indicators: the success probability, defined as the probability of measuring the marked vertex at a given step, and the averaged probability distribution $\overline{P}_T(x,y)$, which captures the long-term spatial behavior of the walk and will be formally introduced below. A numerical approach was necessary, as a full analytical treatment of lackadaisical walks under broken-link decoherence quickly becomes intractable due to the interplay between self-loops and stochastic noise.

\begin{figure}[hbt]
\begin{center}

\includegraphics[width=0.9\columnwidth]{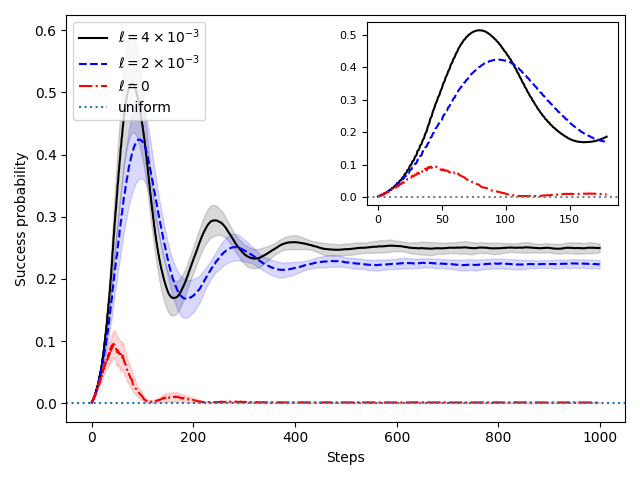}

\end{center}
\caption{\label{fig:prob-vs-steps}Success probability over time under broken-link probability ($p=0.01$) and different self-loop weights, for a $32\times 32$ toroidal grid. Each curve represents the average of 50 independent noisy simulations. The shaded region indicates the standard deviation across runs. The dotted horizontal line marks the uniform distribution for reference. Inset shows early-time behavior. (Color online.)}
\end{figure}

In Fig.~\ref{fig:prob-vs-steps}, we show the time evolution of the success probability
for a $32\times32$ grid under broken-link decoherence with $p=0.01$, comparing
different self-loop weights~$\ell$. Notice that the value $\ell = 4\times10^{-3}$
corresponds approximately to the theoretically motivated optimum $\ell = 4/N$ for
$N=32\times32$. The results show that introducing self-loops systematically
improves the resilience of the search algorithm to noise, yielding higher success
probabilities than in the loopless case. This enhancement is observed for all
values of $\ell$ tested, with the choice $\ell=4/N$ providing the largest and most
stable success probability among the cases shown, while excessively large
self-loop weights eventually lead to reduced performance compared to the optimal scaling. We further observe that
the success probability stabilizes at long times, a behavior that persists for
longer evolutions and higher noise rates.

In all noisy simulations performed, we observe that when the self-loop weight is set to $\ell=4/N$, the long-time success probability approaches a value close to $0.25$. This behavior is consistent across different noise rates and grid sizes, although the approach to the asymptotic regime becomes progressively slower as either $p$ increases or the lattice size grows, requiring long evolution times to clearly resolve the plateau. The existence of this long-time localization is robust in all cases studied, while the
precise asymptotic value is inferred numerically. We further conjecture that this value is related to the degree of the underlying graph, since the two-dimensional grid has degree four, suggesting a limiting probability of order $1/d$ for $d$-regular lattices.

\begin{figure}[hbt]
\begin{center}
\includegraphics[width=0.98\columnwidth]{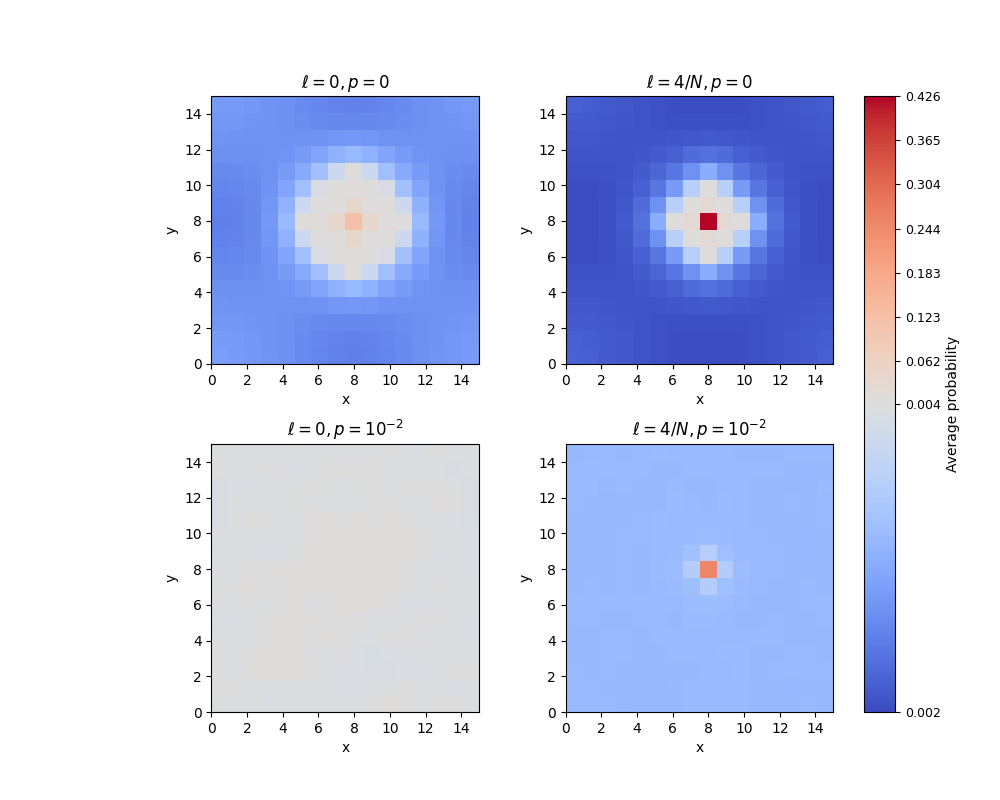}
\end{center}
\caption{\label{fig:heatmaps}Average probability distribution $\overline{P}_T(x,y)$ on a $16\times16$ toroidal grid after $T=10000$ steps of the lackadaisical quantum walk search. Each subplot corresponds to a different pair of parameters, with the broken-link probability $p$ and the self-loop weight $\ell$ indicated in the title; in
particular, the case $\ell=4/N$ corresponds to the value that is optimal in the noiseless setting. The distributions are shown as heatmaps over the grid. (Color online.)
}
\end{figure}

The average probability distribution of the quantum walk is defined as
\begin{equation}
\overline{P}_T(x,y) = \frac{1}{T}\sum_{t=1}^T P_t(x,y),
\end{equation}
where $P_t(x,y)$ is the probability of finding the walker at vertex $(x,y)$ after $t$ steps. In Fig.~\ref{fig:heatmaps}, we have $\overline{P}_T(x,y)$ on a $16\times16$ toroidal grid after $T=10000$ steps, displayed as heatmaps for different values of the broken-link probability~$p$ and the self-loop weight~$\ell$. The upper-left subplot corresponds to the regular algorithm without noise and without self-loops: the marked vertex is identifiable, but only with a small probability of success. In the upper-right subplot, self-loops are included but no noise is present, reproducing Wong’s result~\cite{wong_grover_2015} with a much larger probability at the marked vertex. The bottom-left subplot shows that when noise is introduced without self-loops, decoherence degrades the search completely and the distribution converges to uniform. Finally, the bottom-right subplot illustrates our main finding: using the same self-loop weight $\ell=4/N$ that is optimal in the noiseless case, the marked vertex still stands out with a probability significantly above uniform even in the presence of noise, demonstrating that self-loops enhance the robustness of lackadaisical quantum search under decoherence.

In the noiseless case, the optimal scaling 
$\ell \approx 4/N$ leading to constant success probability on the two-dimensional grid was identified numerically by Wong~\cite{wong_faster_2018}. In the noiseless setting, a general scaling of the optimal self-loop weight for single-target search on vertex-transitive graphs was proposed based on extensive numerical evidence~\cite{rhodes_search_2020}. 

A natural question is whether the optimal self-loop weight $\ell$ must be re-tuned in the presence of broken-link noise, or whether the value that is known to be optimal in the ideal walk ($\ell \approx 4/N$) remains adequate under decoherence.
Before addressing this question quantitatively, notice that Fig.~\ref{fig:heatmaps} provides a
spatially resolved illustration of how the lackadaisical walk with $\ell=4/N$, which is optimal in the noiseless case, responds to broken-link decoherence.
To investigate this further, we performed a sweep over $\ell$ values for several noise rates $p$ and, for each configuration, extracted the maximum success probability achieved during the evolution. 

\begin{figure}
    \centering \includegraphics[width=0.9\linewidth]{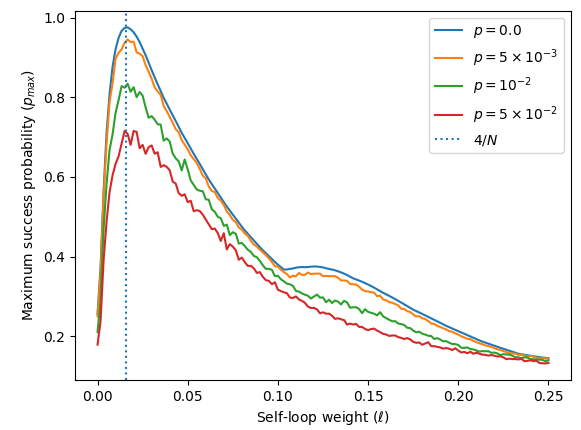}
    \caption{Maximum success probability $P_{\max}$ (maximized over time) as a function of the self-loop weight $\ell$ for several broken-link noise rates $p$ on a $16\times 16$ toroidal grid. 
For all noise levels, the peak remains near the noiseless optimal value $\ell \approx 4/N$, indicating that the parameter 
that maximizes performance is remarkably stable under dynamic percolation noise. Although the overall success probability 
decreases with increasing $p$, the location of the optimal $\ell$ is essentially unchanged, supporting the interpretation that 
the interference structure determining the optimal loop weight in the ideal walk is largely preserved in the noisy regime.}

    \label{fig:maxprob-vs-loop}
\end{figure}

As we can see in Fig.~\ref{fig:maxprob-vs-loop}, the curves for all noise levels peak at approximately the same value $\ell \approx 4/N$, although the absolute success probability decreases with increasing $p$. We considered a $16 \times 16$ grid and took the average over 40 independent noisy simulations. This indicates that the introduction of dynamic percolation does not substantially shift the value of $\ell$ that maximizes performance. The observation suggests that the interference mechanism responsible for the optimal tuning in the ideal walk is largely preserved under moderate structural noise. From a practical perspective, this property is advantageous, as it implies that a single, theoretically motivated choice of $\ell$ remains close to optimal without requiring recalibration under different noise conditions.

\section{Conclusions and discussions}
\label{sec:conclusions}

Quantum walks are a central framework for the design of quantum algorithms, and their practical relevance depends critically on understanding how they behave under realistic imperfections. In this work, we investigated the interplay between self-loops and broken-link decoherence in lackadaisical quantum walks, focusing on their performance in spatial search. Our results show that self-loops do more than optimize the noiseless case: they also mitigate the detrimental effects of decoherence, allowing the marked vertex to remain identifiable even when noise drives the loopless walk toward a uniform distribution. By establishing this robustness, we highlight the strategic role of self-loops as a resource for designing quantum-walk-based search algorithms that remain effective in noisy environments.

We also observe that, for the noiselessly optimal choice $\ell=4/N$, the success probability under broken-link decoherence saturates at a value close to $0.25$ on the two-dimensional grid. While this value is inferred numerically, the persistence of a finite long-time success probability is robust across all cases studied. We conjecture that this plateau is related to the degree of the underlying lattice,
suggesting a limiting probability of order $1/d$ for $d$-regular graphs.

Our results show that, although decoherence inevitably reduces the efficiency of quantum search, the inclusion of self-loops preserves a clear advantage over the loopless case. In particular, while Wong's seminal work~\cite{wong_grover_2015} established the benefit of lackadaisical quantum walks in the noiseless setting, we demonstrate that the same mechanism also enhances robustness under broken-link decoherence, ensuring that the marked vertex still stands out with probability significantly above uniform. Overall, our findings indicate that self-loops are not only a tool for optimizing lackadaisical quantum walks in the ideal case, but also a key ingredient for making quantum search algorithms more resilient to realistic sources of noise. 

A particularly noteworthy observation is that the optimal self-loop weight remains close to the theoretical 
noiseless value $\ell \approx 4/N$ across all noise levels considered. The fact that dynamic edge failures 
reduce overall success probability but do not shift the optimal choice of $\ell$ suggests that the mechanism 
governing constructive interference is remarkably stable. From a practical standpoint, this means that a single, 
theoretically motivated loop parameter can be used to achieve near-optimal performance without the need for 
recalibration in the presence of moderate decoherence.

Although our analysis provides clear evidence of the advantages of self-loops under broken-link decoherence, it remains focused on a specific setting. We considered the two-dimensional toroidal grid and a single type of noise model, which naturally raises the question of how these findings extend to other graphs and decoherence mechanisms. In addition, our study was restricted to the discrete-time lackadaisical walk, leaving open the possibility of exploring whether similar benefits arise in other quantum-walk formulations.

Future work can proceed along several directions. One natural extension is to analyze other noise models beyond broken-link decoherence, such as Gaussian potentials~\cite{de_l_marquezino_impact_2025}, phase damping or depolarizing channels, while connecting to earlier studies of decoherence in quantum search~\cite{abal_decoherence_2009} as a baseline for comparison. Another promising line is to investigate the performance of lackadaisical quantum walks on different graph families, ranging from non-regular, random, or oriented graphs to more applied structures such as Johnson graphs, which could provide insights into noise mitigation strategies for the element distinctness algorithm. 
Another natural extension concerns the case of multiple marked vertices, including clustered configurations on two-dimensional grids, where lackadaisical quantum walks have been shown to achieve near-unit success probability with appropriately tuned self-loop weights in the noiseless setting~\cite{saha_faster_2022, de_carvalho_applying_2023}.
Finally, it would be valuable to explore the role of self-loops in other formulations of quantum walks, such as the staggered model~\cite{santos_decoherence_2022} and continuous-time quantum walks, the latter being directly connected to quantum walk–based optimization approaches like QWOA, thereby broadening the applicability of our findings. 

In summary, our results suggest that self-loops can play a central role in making quantum-walk-based search more resilient to noise. Extending these ideas to other noise models, graph structures, and formulations of quantum walks may further consolidate lackadaisical quantum walks as a versatile tool for robust quantum algorithms.

\backmatter

\bmhead{Acknowledgements}

The authors thank R.~Portugal and L.~Simonetti for helpful discussions.
The authors also thank LabCFD, the Computer Fluid Dynamics laboratory of the Department of Chemical Engineering of UFRJ, for providing access to its computing cluster used in numerical simulations.

\section*{Declarations}


\paragraph{Funding} F.L.M. was partially funded by CNPq Universal Grant process No. 
407296/2021-2, and
CNPq Productivity Grant process No. 306049/2025-2.
The authors acknowledge the financial support of the National Institute of Science and Technology for Applied Quantum Computing through CNPq process No. 408884/2024-0.

\paragraph{Competing interests} The authors declare no competing interests.

\paragraph{Ethics approval} Not applicable.

\paragraph{Data availability and code availability}

The data used in this study can be generated with the computer code available at \url{https://github.com/Marquezino/NoiseLackadaisicalQW}.

\paragraph{Author contribution}

The authors' contributions to this study are as follows.
\textbf{G. M. O. Vieira}: Investigation, Software, Writing -- review \& editing, Visualization;
\textbf{N. Maculan}: Investigation, Supervision, Writing -- review \& editing;
\textbf{F. L. Marquezino}: Investigation, Supervision, Conceptualization, Methodology, Software, Writing -- original draft, Visualization.

\bibliography{gabriel-artigo-QINP}

\end{document}

%% file: lackadaisical.tikz
\tikzset{every picture/.style={line width=0.75pt}} 

\begin{tikzpicture}[x=0.75pt,y=0.75pt,yscale=-1,xscale=1]

\draw  [draw opacity=0][line width=1.5]  (90,84) -- (314,84) -- (314,308) -- (90,308) -- cycle ; \draw  [line width=1.5]  (111,84) -- (111,308)(171,84) -- (171,308)(231,84) -- (231,308)(291,84) -- (291,308) ; \draw  [line width=1.5]  (90,105) -- (314,105)(90,165) -- (314,165)(90,225) -- (314,225)(90,285) -- (314,285) ; \draw  [line width=1.5]   ;
\draw  [fill={rgb, 255:red, 0; green, 0; blue, 0 }  ,fill opacity=1 ][line width=1.5]  (282,105) .. controls (282,100.03) and (286.03,96) .. (291,96) .. controls (295.97,96) and (300,100.03) .. (300,105) .. controls (300,109.97) and (295.97,114) .. (291,114) .. controls (286.03,114) and (282,109.97) .. (282,105) -- cycle ;
\draw  [fill={rgb, 255:red, 0; green, 0; blue, 0 }  ,fill opacity=1 ][line width=1.5]  (222,105) .. controls (222,100.03) and (226.03,96) .. (231,96) .. controls (235.97,96) and (240,100.03) .. (240,105) .. controls (240,109.97) and (235.97,114) .. (231,114) .. controls (226.03,114) and (222,109.97) .. (222,105) -- cycle ;
\draw  [fill={rgb, 255:red, 0; green, 0; blue, 0 }  ,fill opacity=1 ][line width=1.5]  (162,105) .. controls (162,100.03) and (166.03,96) .. (171,96) .. controls (175.97,96) and (180,100.03) .. (180,105) .. controls (180,109.97) and (175.97,114) .. (171,114) .. controls (166.03,114) and (162,109.97) .. (162,105) -- cycle ;
\draw  [fill={rgb, 255:red, 0; green, 0; blue, 0 }  ,fill opacity=1 ][line width=1.5]  (102,105) .. controls (102,100.03) and (106.03,96) .. (111,96) .. controls (115.97,96) and (120,100.03) .. (120,105) .. controls (120,109.97) and (115.97,114) .. (111,114) .. controls (106.03,114) and (102,109.97) .. (102,105) -- cycle ;
\draw  [fill={rgb, 255:red, 0; green, 0; blue, 0 }  ,fill opacity=1 ][line width=1.5]  (102,165) .. controls (102,160.03) and (106.03,156) .. (111,156) .. controls (115.97,156) and (120,160.03) .. (120,165) .. controls (120,169.97) and (115.97,174) .. (111,174) .. controls (106.03,174) and (102,169.97) .. (102,165) -- cycle ;
\draw  [fill={rgb, 255:red, 0; green, 0; blue, 0 }  ,fill opacity=1 ][line width=1.5]  (162,165) .. controls (162,160.03) and (166.03,156) .. (171,156) .. controls (175.97,156) and (180,160.03) .. (180,165) .. controls (180,169.97) and (175.97,174) .. (171,174) .. controls (166.03,174) and (162,169.97) .. (162,165) -- cycle ;
\draw  [fill={rgb, 255:red, 255; green, 255; blue, 255 }  ,fill opacity=1 ][line width=3]  (222,165) .. controls (222,160.03) and (226.03,156) .. (231,156) .. controls (235.97,156) and (240,160.03) .. (240,165) .. controls (240,169.97) and (235.97,174) .. (231,174) .. controls (226.03,174) and (222,169.97) .. (222,165) -- cycle ;
\draw  [fill={rgb, 255:red, 0; green, 0; blue, 0 }  ,fill opacity=1 ][line width=1.5]  (282,165) .. controls (282,160.03) and (286.03,156) .. (291,156) .. controls (295.97,156) and (300,160.03) .. (300,165) .. controls (300,169.97) and (295.97,174) .. (291,174) .. controls (286.03,174) and (282,169.97) .. (282,165) -- cycle ;
\draw  [fill={rgb, 255:red, 0; green, 0; blue, 0 }  ,fill opacity=1 ][line width=1.5]  (102,225) .. controls (102,220.03) and (106.03,216) .. (111,216) .. controls (115.97,216) and (120,220.03) .. (120,225) .. controls (120,229.97) and (115.97,234) .. (111,234) .. controls (106.03,234) and (102,229.97) .. (102,225) -- cycle ;
\draw  [fill={rgb, 255:red, 0; green, 0; blue, 0 }  ,fill opacity=1 ][line width=1.5]  (162,225) .. controls (162,220.03) and (166.03,216) .. (171,216) .. controls (175.97,216) and (180,220.03) .. (180,225) .. controls (180,229.97) and (175.97,234) .. (171,234) .. controls (166.03,234) and (162,229.97) .. (162,225) -- cycle ;
\draw  [fill={rgb, 255:red, 0; green, 0; blue, 0 }  ,fill opacity=1 ][line width=1.5]  (222,225) .. controls (222,220.03) and (226.03,216) .. (231,216) .. controls (235.97,216) and (240,220.03) .. (240,225) .. controls (240,229.97) and (235.97,234) .. (231,234) .. controls (226.03,234) and (222,229.97) .. (222,225) -- cycle ;
\draw  [fill={rgb, 255:red, 0; green, 0; blue, 0 }  ,fill opacity=1 ][line width=1.5]  (282,225) .. controls (282,220.03) and (286.03,216) .. (291,216) .. controls (295.97,216) and (300,220.03) .. (300,225) .. controls (300,229.97) and (295.97,234) .. (291,234) .. controls (286.03,234) and (282,229.97) .. (282,225) -- cycle ;
\draw  [fill={rgb, 255:red, 0; green, 0; blue, 0 }  ,fill opacity=1 ][line width=1.5]  (102,285) .. controls (102,280.03) and (106.03,276) .. (111,276) .. controls (115.97,276) and (120,280.03) .. (120,285) .. controls (120,289.97) and (115.97,294) .. (111,294) .. controls (106.03,294) and (102,289.97) .. (102,285) -- cycle ;
\draw  [fill={rgb, 255:red, 0; green, 0; blue, 0 }  ,fill opacity=1 ][line width=1.5]  (162,285) .. controls (162,280.03) and (166.03,276) .. (171,276) .. controls (175.97,276) and (180,280.03) .. (180,285) .. controls (180,289.97) and (175.97,294) .. (171,294) .. controls (166.03,294) and (162,289.97) .. (162,285) -- cycle ;
\draw  [fill={rgb, 255:red, 0; green, 0; blue, 0 }  ,fill opacity=1 ][line width=1.5]  (222,285) .. controls (222,280.03) and (226.03,276) .. (231,276) .. controls (235.97,276) and (240,280.03) .. (240,285) .. controls (240,289.97) and (235.97,294) .. (231,294) .. controls (226.03,294) and (222,289.97) .. (222,285) -- cycle ;
\draw  [fill={rgb, 255:red, 0; green, 0; blue, 0 }  ,fill opacity=1 ][line width=1.5]  (282,285) .. controls (282,280.03) and (286.03,276) .. (291,276) .. controls (295.97,276) and (300,280.03) .. (300,285) .. controls (300,289.97) and (295.97,294) .. (291,294) .. controls (286.03,294) and (282,289.97) .. (282,285) -- cycle ;
\draw [line width=1.5]  [dash pattern={on 1.69pt off 2.76pt}]  (300,105) -- (340,105) ;
\draw [line width=1.5]  [dash pattern={on 1.69pt off 2.76pt}]  (300,165) -- (340,165) ;
\draw [line width=1.5]  [dash pattern={on 1.69pt off 2.76pt}]  (300,225) -- (340,225) ;
\draw [line width=1.5]  [dash pattern={on 1.69pt off 2.76pt}]  (300,285) -- (340,285) ;
\draw [line width=1.5]  [dash pattern={on 1.69pt off 2.76pt}]  (291,294) -- (291,322) ;
\draw [line width=1.5]  [dash pattern={on 1.69pt off 2.76pt}]  (231,294) -- (231,322) ;
\draw [line width=1.5]  [dash pattern={on 1.69pt off 2.76pt}]  (171,294) -- (171,322) ;
\draw [line width=1.5]  [dash pattern={on 1.69pt off 2.76pt}]  (111,294) -- (111,322) ;
\draw [line width=1.5]    (291,96) .. controls (302,78) and (318,69) .. (323,73) .. controls (327.88,76.9) and (328,88.4) .. (293.71,95.47) ;
\draw [shift={(291,96)}, rotate = 349.29] [color={rgb, 255:red, 0; green, 0; blue, 0 }  ][line width=1.5]    (14.21,-4.28) .. controls (9.04,-1.82) and (4.3,-0.39) .. (0,0) .. controls (4.3,0.39) and (9.04,1.82) .. (14.21,4.28)   ;
\draw [line width=1.5]    (291,156) .. controls (302,138) and (318,129) .. (323,133) .. controls (327.88,136.9) and (328,148.41) .. (293.71,155.47) ;
\draw [shift={(291,156)}, rotate = 349.29] [color={rgb, 255:red, 0; green, 0; blue, 0 }  ][line width=1.5]    (14.21,-4.28) .. controls (9.04,-1.82) and (4.3,-0.39) .. (0,0) .. controls (4.3,0.39) and (9.04,1.82) .. (14.21,4.28)   ;
\draw [line width=1.5]    (291,216) .. controls (302,198) and (318,189) .. (323,193) .. controls (327.88,196.9) and (328,208.41) .. (293.71,215.47) ;
\draw [shift={(291,216)}, rotate = 349.29] [color={rgb, 255:red, 0; green, 0; blue, 0 }  ][line width=1.5]    (14.21,-4.28) .. controls (9.04,-1.82) and (4.3,-0.39) .. (0,0) .. controls (4.3,0.39) and (9.04,1.82) .. (14.21,4.28)   ;
\draw [line width=1.5]    (291,276) .. controls (302,258) and (318,249) .. (323,253) .. controls (327.88,256.9) and (328,268.4) .. (293.71,275.47) ;
\draw [shift={(291,276)}, rotate = 349.29] [color={rgb, 255:red, 0; green, 0; blue, 0 }  ][line width=1.5]    (14.21,-4.28) .. controls (9.04,-1.82) and (4.3,-0.39) .. (0,0) .. controls (4.3,0.39) and (9.04,1.82) .. (14.21,4.28)   ;
\draw [line width=1.5]    (231,96) .. controls (242,78) and (258,69) .. (263,73) .. controls (267.88,76.9) and (268,88.4) .. (233.71,95.47) ;
\draw [shift={(231,96)}, rotate = 349.29] [color={rgb, 255:red, 0; green, 0; blue, 0 }  ][line width=1.5]    (14.21,-4.28) .. controls (9.04,-1.82) and (4.3,-0.39) .. (0,0) .. controls (4.3,0.39) and (9.04,1.82) .. (14.21,4.28)   ;
\draw [line width=1.5]    (231,156) .. controls (242,138) and (258,129) .. (263,133) .. controls (267.88,136.9) and (268,148.41) .. (233.71,155.47) ;
\draw [shift={(231,156)}, rotate = 349.29] [color={rgb, 255:red, 0; green, 0; blue, 0 }  ][line width=1.5]    (14.21,-4.28) .. controls (9.04,-1.82) and (4.3,-0.39) .. (0,0) .. controls (4.3,0.39) and (9.04,1.82) .. (14.21,4.28)   ;
\draw [line width=1.5]    (231,216) .. controls (242,198) and (258,189) .. (263,193) .. controls (267.88,196.9) and (268,208.41) .. (233.71,215.47) ;
\draw [shift={(231,216)}, rotate = 349.29] [color={rgb, 255:red, 0; green, 0; blue, 0 }  ][line width=1.5]    (14.21,-4.28) .. controls (9.04,-1.82) and (4.3,-0.39) .. (0,0) .. controls (4.3,0.39) and (9.04,1.82) .. (14.21,4.28)   ;
\draw [line width=1.5]    (231,276) .. controls (242,258) and (258,249) .. (263,253) .. controls (267.88,256.9) and (268,268.4) .. (233.71,275.47) ;
\draw [shift={(231,276)}, rotate = 349.29] [color={rgb, 255:red, 0; green, 0; blue, 0 }  ][line width=1.5]    (14.21,-4.28) .. controls (9.04,-1.82) and (4.3,-0.39) .. (0,0) .. controls (4.3,0.39) and (9.04,1.82) .. (14.21,4.28)   ;
\draw [line width=1.5]    (171,96) .. controls (182,78) and (198,69) .. (203,73) .. controls (207.88,76.9) and (208,88.4) .. (173.71,95.47) ;
\draw [shift={(171,96)}, rotate = 349.29] [color={rgb, 255:red, 0; green, 0; blue, 0 }  ][line width=1.5]    (14.21,-4.28) .. controls (9.04,-1.82) and (4.3,-0.39) .. (0,0) .. controls (4.3,0.39) and (9.04,1.82) .. (14.21,4.28)   ;
\draw [line width=1.5]    (171,156) .. controls (182,138) and (198,129) .. (203,133) .. controls (207.88,136.9) and (208,148.41) .. (173.71,155.47) ;
\draw [shift={(171,156)}, rotate = 349.29] [color={rgb, 255:red, 0; green, 0; blue, 0 }  ][line width=1.5]    (14.21,-4.28) .. controls (9.04,-1.82) and (4.3,-0.39) .. (0,0) .. controls (4.3,0.39) and (9.04,1.82) .. (14.21,4.28)   ;
\draw [line width=1.5]    (171,216) .. controls (182,198) and (198,189) .. (203,193) .. controls (207.88,196.9) and (208,208.41) .. (173.71,215.47) ;
\draw [shift={(171,216)}, rotate = 349.29] [color={rgb, 255:red, 0; green, 0; blue, 0 }  ][line width=1.5]    (14.21,-4.28) .. controls (9.04,-1.82) and (4.3,-0.39) .. (0,0) .. controls (4.3,0.39) and (9.04,1.82) .. (14.21,4.28)   ;
\draw [line width=1.5]    (171,276) .. controls (182,258) and (198,249) .. (203,253) .. controls (207.88,256.9) and (208,268.4) .. (173.71,275.47) ;
\draw [shift={(171,276)}, rotate = 349.29] [color={rgb, 255:red, 0; green, 0; blue, 0 }  ][line width=1.5]    (14.21,-4.28) .. controls (9.04,-1.82) and (4.3,-0.39) .. (0,0) .. controls (4.3,0.39) and (9.04,1.82) .. (14.21,4.28)   ;
\draw [line width=1.5]    (111,276) .. controls (122,258) and (138,249) .. (143,253) .. controls (147.88,256.9) and (148,268.4) .. (113.71,275.47) ;
\draw [shift={(111,276)}, rotate = 349.29] [color={rgb, 255:red, 0; green, 0; blue, 0 }  ][line width=1.5]    (14.21,-4.28) .. controls (9.04,-1.82) and (4.3,-0.39) .. (0,0) .. controls (4.3,0.39) and (9.04,1.82) .. (14.21,4.28)   ;
\draw [line width=1.5]    (111,216) .. controls (122,198) and (138,189) .. (143,193) .. controls (147.88,196.9) and (148,208.41) .. (113.71,215.47) ;
\draw [shift={(111,216)}, rotate = 349.29] [color={rgb, 255:red, 0; green, 0; blue, 0 }  ][line width=1.5]    (14.21,-4.28) .. controls (9.04,-1.82) and (4.3,-0.39) .. (0,0) .. controls (4.3,0.39) and (9.04,1.82) .. (14.21,4.28)   ;
\draw [line width=1.5]    (111,156) .. controls (122,138) and (138,129) .. (143,133) .. controls (147.88,136.9) and (148,148.41) .. (113.71,155.47) ;
\draw [shift={(111,156)}, rotate = 349.29] [color={rgb, 255:red, 0; green, 0; blue, 0 }  ][line width=1.5]    (14.21,-4.28) .. controls (9.04,-1.82) and (4.3,-0.39) .. (0,0) .. controls (4.3,0.39) and (9.04,1.82) .. (14.21,4.28)   ;
\draw [line width=1.5]    (111,96) .. controls (122,78) and (138,69) .. (143,73) .. controls (147.88,76.9) and (148,88.4) .. (113.71,95.47) ;
\draw [shift={(111,96)}, rotate = 349.29] [color={rgb, 255:red, 0; green, 0; blue, 0 }  ][line width=1.5]    (14.21,-4.28) .. controls (9.04,-1.82) and (4.3,-0.39) .. (0,0) .. controls (4.3,0.39) and (9.04,1.82) .. (14.21,4.28)   ;

\draw (145,54.4) node [anchor=north west][inner sep=0.75pt]  [font=\large]  {$\ell $};

\draw   (289.46,94.46) -- (292.54,97.54)(292.54,94.46) -- (289.46,97.54) ;
\draw   (298.46,103.46) -- (301.54,106.54)(301.54,103.46) -- (298.46,106.54) ;
\draw   (280.46,103.46) -- (283.54,106.54)(283.54,103.46) -- (280.46,106.54) ;
\draw   (229.46,94.46) -- (232.54,97.54)(232.54,94.46) -- (229.46,97.54) ;
\draw   (229.46,94.46) -- (232.54,97.54)(232.54,94.46) -- (229.46,97.54) ;
\draw   (229.46,112.46) -- (232.54,115.54)(232.54,112.46) -- (229.46,115.54) ;
\draw   (229.46,112.46) -- (232.54,115.54)(232.54,112.46) -- (229.46,115.54) ;
\draw   (238.46,103.46) -- (241.54,106.54)(241.54,103.46) -- (238.46,106.54) ;
\draw   (220.46,103.46) -- (223.54,106.54)(223.54,103.46) -- (220.46,106.54) ;
\draw   (169.46,94.46) -- (172.54,97.54)(172.54,94.46) -- (169.46,97.54) ;
\draw   (169.46,94.46) -- (172.54,97.54)(172.54,94.46) -- (169.46,97.54) ;
\draw   (169.46,112.46) -- (172.54,115.54)(172.54,112.46) -- (169.46,115.54) ;
\draw   (169.46,112.46) -- (172.54,115.54)(172.54,112.46) -- (169.46,115.54) ;
\draw   (178.46,103.46) -- (181.54,106.54)(181.54,103.46) -- (178.46,106.54) ;
\draw   (160.46,103.46) -- (163.54,106.54)(163.54,103.46) -- (160.46,106.54) ;
\draw   (109.46,94.46) -- (112.54,97.54)(112.54,94.46) -- (109.46,97.54) ;
\draw   (109.46,112.46) -- (112.54,115.54)(112.54,112.46) -- (109.46,115.54) ;
\draw   (118.46,103.46) -- (121.54,106.54)(121.54,103.46) -- (118.46,106.54) ;
\draw   (100.46,103.46) -- (103.54,106.54)(103.54,103.46) -- (100.46,106.54) ;
\draw   (109.46,154.46) -- (112.54,157.54)(112.54,154.46) -- (109.46,157.54) ;
\draw   (109.46,172.46) -- (112.54,175.54)(112.54,172.46) -- (109.46,175.54) ;
\draw   (100.46,163.46) -- (103.54,166.54)(103.54,163.46) -- (100.46,166.54) ;
\draw   (118.46,163.46) -- (121.54,166.54)(121.54,163.46) -- (118.46,166.54) ;
\draw   (100.46,163.46) -- (103.54,166.54)(103.54,163.46) -- (100.46,166.54) ;
\draw   (169.46,154.46) -- (172.54,157.54)(172.54,154.46) -- (169.46,157.54) ;
\draw   (169.46,154.46) -- (172.54,157.54)(172.54,154.46) -- (169.46,157.54) ;
\draw   (169.46,172.46) -- (172.54,175.54)(172.54,172.46) -- (169.46,175.54) ;
\draw   (169.46,172.46) -- (172.54,175.54)(172.54,172.46) -- (169.46,175.54) ;
\draw   (160.46,163.46) -- (163.54,166.54)(163.54,163.46) -- (160.46,166.54) ;
\draw   (178.46,163.46) -- (181.54,166.54)(181.54,163.46) -- (178.46,166.54) ;
\draw   (160.46,163.46) -- (163.54,166.54)(163.54,163.46) -- (160.46,166.54) ;
\draw   (229.46,154.46) -- (232.54,157.54)(232.54,154.46) -- (229.46,157.54) ;
\draw   (229.46,154.46) -- (232.54,157.54)(232.54,154.46) -- (229.46,157.54) ;
\draw   (229.46,172.46) -- (232.54,175.54)(232.54,172.46) -- (229.46,175.54) ;
\draw   (229.46,172.46) -- (232.54,175.54)(232.54,172.46) -- (229.46,175.54) ;
\draw   (220.46,163.46) -- (223.54,166.54)(223.54,163.46) -- (220.46,166.54) ;
\draw   (238.46,163.46) -- (241.54,166.54)(241.54,163.46) -- (238.46,166.54) ;
\draw   (220.46,163.46) -- (223.54,166.54)(223.54,163.46) -- (220.46,166.54) ;
\draw   (289.46,154.46) -- (292.54,157.54)(292.54,154.46) -- (289.46,157.54) ;
\draw   (280.46,163.46) -- (283.54,166.54)(283.54,163.46) -- (280.46,166.54) ;
\draw   (298.46,163.46) -- (301.54,166.54)(301.54,163.46) -- (298.46,166.54) ;
\draw   (280.46,163.46) -- (283.54,166.54)(283.54,163.46) -- (280.46,166.54) ;
\draw   (109.46,214.46) -- (112.54,217.54)(112.54,214.46) -- (109.46,217.54) ;
\draw   (109.46,232.46) -- (112.54,235.54)(112.54,232.46) -- (109.46,235.54) ;
\draw   (100.46,223.46) -- (103.54,226.54)(103.54,223.46) -- (100.46,226.54) ;
\draw   (118.46,223.46) -- (121.54,226.54)(121.54,223.46) -- (118.46,226.54) ;
\draw   (169.46,214.46) -- (172.54,217.54)(172.54,214.46) -- (169.46,217.54) ;
\draw   (169.46,214.46) -- (172.54,217.54)(172.54,214.46) -- (169.46,217.54) ;
\draw   (169.46,232.46) -- (172.54,235.54)(172.54,232.46) -- (169.46,235.54) ;
\draw   (169.46,232.46) -- (172.54,235.54)(172.54,232.46) -- (169.46,235.54) ;
\draw   (160.46,223.46) -- (163.54,226.54)(163.54,223.46) -- (160.46,226.54) ;
\draw   (178.46,223.46) -- (181.54,226.54)(181.54,223.46) -- (178.46,226.54) ;
\draw   (229.46,214.46) -- (232.54,217.54)(232.54,214.46) -- (229.46,217.54) ;
\draw   (229.46,214.46) -- (232.54,217.54)(232.54,214.46) -- (229.46,217.54) ;
\draw   (229.46,232.46) -- (232.54,235.54)(232.54,232.46) -- (229.46,235.54) ;
\draw   (229.46,232.46) -- (232.54,235.54)(232.54,232.46) -- (229.46,235.54) ;
\draw   (220.46,223.46) -- (223.54,226.54)(223.54,223.46) -- (220.46,226.54) ;
\draw   (238.46,223.46) -- (241.54,226.54)(241.54,223.46) -- (238.46,226.54) ;
\draw   (289.46,214.46) -- (292.54,217.54)(292.54,214.46) -- (289.46,217.54) ;
\draw   (280.46,223.46) -- (283.54,226.54)(283.54,223.46) -- (280.46,226.54) ;
\draw   (298.46,223.46) -- (301.54,226.54)(301.54,223.46) -- (298.46,226.54) ;
\draw   (109.46,274.46) -- (112.54,277.54)(112.54,274.46) -- (109.46,277.54) ;
\draw   (109.46,292.46) -- (112.54,295.54)(112.54,292.46) -- (109.46,295.54) ;
\draw   (118.46,283.46) -- (121.54,286.54)(121.54,283.46) -- (118.46,286.54) ;
\draw   (100.46,283.46) -- (103.54,286.54)(103.54,283.46) -- (100.46,286.54) ;
\draw   (169.46,274.46) -- (172.54,277.54)(172.54,274.46) -- (169.46,277.54) ;
\draw   (169.46,274.46) -- (172.54,277.54)(172.54,274.46) -- (169.46,277.54) ;
\draw   (169.46,292.46) -- (172.54,295.54)(172.54,292.46) -- (169.46,295.54) ;
\draw   (169.46,292.46) -- (172.54,295.54)(172.54,292.46) -- (169.46,295.54) ;
\draw   (178.46,283.46) -- (181.54,286.54)(181.54,283.46) -- (178.46,286.54) ;
\draw   (160.46,283.46) -- (163.54,286.54)(163.54,283.46) -- (160.46,286.54) ;
\draw   (229.46,274.46) -- (232.54,277.54)(232.54,274.46) -- (229.46,277.54) ;
\draw   (229.46,274.46) -- (232.54,277.54)(232.54,274.46) -- (229.46,277.54) ;
\draw   (229.46,292.46) -- (232.54,295.54)(232.54,292.46) -- (229.46,295.54) ;
\draw   (229.46,292.46) -- (232.54,295.54)(232.54,292.46) -- (229.46,295.54) ;
\draw   (238.46,283.46) -- (241.54,286.54)(241.54,283.46) -- (238.46,286.54) ;
\draw   (220.46,283.46) -- (223.54,286.54)(223.54,283.46) -- (220.46,286.54) ;
\draw   (289.46,274.46) -- (292.54,277.54)(292.54,274.46) -- (289.46,277.54) ;
\draw   (298.46,283.46) -- (301.54,286.54)(301.54,283.46) -- (298.46,286.54) ;
\draw   (280.46,283.46) -- (283.54,286.54)(283.54,283.46) -- (280.46,286.54) ;
\draw   (289.46,94.46) -- (292.54,97.54)(292.54,94.46) -- (289.46,97.54) ;
\draw   (289.46,154.46) -- (292.54,157.54)(292.54,154.46) -- (289.46,157.54) ;
\draw   (289.46,214.46) -- (292.54,217.54)(292.54,214.46) -- (289.46,217.54) ;
\draw   (289.46,274.46) -- (292.54,277.54)(292.54,274.46) -- (289.46,277.54) ;
\draw   (229.46,94.46) -- (232.54,97.54)(232.54,94.46) -- (229.46,97.54) ;
\draw   (229.46,94.46) -- (232.54,97.54)(232.54,94.46) -- (229.46,97.54) ;
\draw   (229.46,154.46) -- (232.54,157.54)(232.54,154.46) -- (229.46,157.54) ;
\draw   (229.46,154.46) -- (232.54,157.54)(232.54,154.46) -- (229.46,157.54) ;
\draw   (229.46,214.46) -- (232.54,217.54)(232.54,214.46) -- (229.46,217.54) ;
\draw   (229.46,214.46) -- (232.54,217.54)(232.54,214.46) -- (229.46,217.54) ;
\draw   (229.46,274.46) -- (232.54,277.54)(232.54,274.46) -- (229.46,277.54) ;
\draw   (229.46,274.46) -- (232.54,277.54)(232.54,274.46) -- (229.46,277.54) ;
\draw   (169.46,94.46) -- (172.54,97.54)(172.54,94.46) -- (169.46,97.54) ;
\draw   (169.46,94.46) -- (172.54,97.54)(172.54,94.46) -- (169.46,97.54) ;
\draw   (169.46,154.46) -- (172.54,157.54)(172.54,154.46) -- (169.46,157.54) ;
\draw   (169.46,154.46) -- (172.54,157.54)(172.54,154.46) -- (169.46,157.54) ;
\draw   (169.46,214.46) -- (172.54,217.54)(172.54,214.46) -- (169.46,217.54) ;
\draw   (169.46,214.46) -- (172.54,217.54)(172.54,214.46) -- (169.46,217.54) ;
\draw   (169.46,274.46) -- (172.54,277.54)(172.54,274.46) -- (169.46,277.54) ;
\draw   (169.46,274.46) -- (172.54,277.54)(172.54,274.46) -- (169.46,277.54) ;
\draw   (109.46,274.46) -- (112.54,277.54)(112.54,274.46) -- (109.46,277.54) ;
\draw   (109.46,274.46) -- (112.54,277.54)(112.54,274.46) -- (109.46,277.54) ;
\draw   (109.46,214.46) -- (112.54,217.54)(112.54,214.46) -- (109.46,217.54) ;
\draw   (109.46,214.46) -- (112.54,217.54)(112.54,214.46) -- (109.46,217.54) ;
\draw   (109.46,154.46) -- (112.54,157.54)(112.54,154.46) -- (109.46,157.54) ;
\draw   (109.46,94.46) -- (112.54,97.54)(112.54,94.46) -- (109.46,97.54) ;
\draw   (109.46,94.46) -- (112.54,97.54)(112.54,94.46) -- (109.46,97.54) ;
\end{tikzpicture}

%% file: broken-links.tikz
\tikzset{every picture/.style={line width=0.75pt}} 

\begin{tikzpicture}
[x=0.75pt,y=0.75pt,yscale=-0.8,xscale=0.8]

\draw  [fill={rgb, 255:red, 0; green, 0; blue, 0 }  ,fill opacity=1 ][line width=1.5]  (149,84.5) .. controls (149,79.53) and (153.03,75.5) .. (158,75.5) .. controls (162.97,75.5) and (167,79.53) .. (167,84.5) .. controls (167,89.47) and (162.97,93.5) .. (158,93.5) .. controls (153.03,93.5) and (149,89.47) .. (149,84.5) -- cycle ;
\draw  [fill={rgb, 255:red, 0; green, 0; blue, 0 }  ,fill opacity=1 ][line width=1.5]  (232.75,84.5) .. controls (232.75,79.53) and (236.78,75.5) .. (241.75,75.5) .. controls (246.72,75.5) and (250.75,79.53) .. (250.75,84.5) .. controls (250.75,89.47) and (246.72,93.5) .. (241.75,93.5) .. controls (236.78,93.5) and (232.75,89.47) .. (232.75,84.5) -- cycle ;
\draw  [fill={rgb, 255:red, 0; green, 0; blue, 0 }  ,fill opacity=1 ][line width=1.5]  (316.5,84.5) .. controls (316.5,79.53) and (320.53,75.5) .. (325.5,75.5) .. controls (330.47,75.5) and (334.5,79.53) .. (334.5,84.5) .. controls (334.5,89.47) and (330.47,93.5) .. (325.5,93.5) .. controls (320.53,93.5) and (316.5,89.47) .. (316.5,84.5) -- cycle ;
\draw  [fill={rgb, 255:red, 0; green, 0; blue, 0 }  ,fill opacity=1 ][line width=1.5]  (400.25,84.5) .. controls (400.25,79.53) and (404.28,75.5) .. (409.25,75.5) .. controls (414.22,75.5) and (418.25,79.53) .. (418.25,84.5) .. controls (418.25,89.47) and (414.22,93.5) .. (409.25,93.5) .. controls (404.28,93.5) and (400.25,89.47) .. (400.25,84.5) -- cycle ;
\draw  [fill={rgb, 255:red, 0; green, 0; blue, 0 }  ,fill opacity=1 ][line width=1.5]  (484,84.5) .. controls (484,79.53) and (488.03,75.5) .. (493,75.5) .. controls (497.97,75.5) and (502,79.53) .. (502,84.5) .. controls (502,89.47) and (497.97,93.5) .. (493,93.5) .. controls (488.03,93.5) and (484,89.47) .. (484,84.5) -- cycle ;
\draw [line width=1.5]    (158,75.5) .. controls (177.5,46.74) and (213.4,40.79) .. (235.34,69.71) ;
\draw [shift={(237,72)}, rotate = 235.38] [color={rgb, 255:red, 0; green, 0; blue, 0 }  ][line width=1.5]    (14.21,-6.37) .. controls (9.04,-2.99) and (4.3,-0.87) .. (0,0) .. controls (4.3,0.87) and (9.04,2.99) .. (14.21,6.37)   ;
\draw [line width=1.5]    (241.75,75.5) .. controls (261.25,46.74) and (298.34,39.84) .. (320.34,68.71) ;
\draw [shift={(322,71)}, rotate = 235.38] [color={rgb, 255:red, 0; green, 0; blue, 0 }  ][line width=1.5]    (14.21,-6.37) .. controls (9.04,-2.99) and (4.3,-0.87) .. (0,0) .. controls (4.3,0.87) and (9.04,2.99) .. (14.21,6.37)   ;
\draw [line width=1.5]    (325.5,75.5) .. controls (345,46.74) and (380.42,39.84) .. (402.34,68.71) ;
\draw [shift={(404,71)}, rotate = 235.38] [color={rgb, 255:red, 0; green, 0; blue, 0 }  ][line width=1.5]    (14.21,-6.37) .. controls (9.04,-2.99) and (4.3,-0.87) .. (0,0) .. controls (4.3,0.87) and (9.04,2.99) .. (14.21,6.37)   ;
\draw [line width=1.5]    (409.25,75.5) .. controls (428.75,46.74) and (466.31,39.84) .. (488.34,68.71) ;
\draw [shift={(490,71)}, rotate = 235.38] [color={rgb, 255:red, 0; green, 0; blue, 0 }  ][line width=1.5]    (14.21,-6.37) .. controls (9.04,-2.99) and (4.3,-0.87) .. (0,0) .. controls (4.3,0.87) and (9.04,2.99) .. (14.21,6.37)   ;
\draw [line width=1.5]    (164.76,100.49) .. controls (184.27,126.56) and (220.65,126.97) .. (241.75,93.5) ;
\draw [shift={(163,98)}, rotate = 56.31] [color={rgb, 255:red, 0; green, 0; blue, 0 }  ][line width=1.5]    (14.21,-6.37) .. controls (9.04,-2.99) and (4.3,-0.87) .. (0,0) .. controls (4.3,0.87) and (9.04,2.99) .. (14.21,6.37)   ;
\draw [line width=1.5]    (247.76,101.49) .. controls (267.31,127.5) and (304.4,126.97) .. (325.5,93.5) ;
\draw [shift={(246,99)}, rotate = 56.31] [color={rgb, 255:red, 0; green, 0; blue, 0 }  ][line width=1.5]    (14.21,-6.37) .. controls (9.04,-2.99) and (4.3,-0.87) .. (0,0) .. controls (4.3,0.87) and (9.04,2.99) .. (14.21,6.37)   ;
\draw [line width=1.5]    (331.76,100.49) .. controls (351.29,126.56) and (388.15,126.97) .. (409.25,93.5) ;
\draw [shift={(330,98)}, rotate = 56.31] [color={rgb, 255:red, 0; green, 0; blue, 0 }  ][line width=1.5]    (14.21,-6.37) .. controls (9.04,-2.99) and (4.3,-0.87) .. (0,0) .. controls (4.3,0.87) and (9.04,2.99) .. (14.21,6.37)   ;
\draw [line width=1.5]    (414.76,100.49) .. controls (434.34,126.56) and (471.9,126.97) .. (493,93.5) ;
\draw [shift={(413,98)}, rotate = 56.31] [color={rgb, 255:red, 0; green, 0; blue, 0 }  ][line width=1.5]    (14.21,-6.37) .. controls (9.04,-2.99) and (4.3,-0.87) .. (0,0) .. controls (4.3,0.87) and (9.04,2.99) .. (14.21,6.37)   ;

\draw  [fill={rgb, 255:red, 0; green, 0; blue, 0 }  ,fill opacity=1 ][line width=1.5]  (149,219.5) .. controls (149,214.53) and (153.03,210.5) .. (158,210.5) .. controls (162.97,210.5) and (167,214.53) .. (167,219.5) .. controls (167,224.47) and (162.97,228.5) .. (158,228.5) .. controls (153.03,228.5) and (149,224.47) .. (149,219.5) -- cycle ;
\draw  [fill={rgb, 255:red, 0; green, 0; blue, 0 }  ,fill opacity=1 ][line width=1.5]  (232.75,219.5) .. controls (232.75,214.53) and (236.78,210.5) .. (241.75,210.5) .. controls (246.72,210.5) and (250.75,214.53) .. (250.75,219.5) .. controls (250.75,224.47) and (246.72,228.5) .. (241.75,228.5) .. controls (236.78,228.5) and (232.75,224.47) .. (232.75,219.5) -- cycle ;
\draw  [fill={rgb, 255:red, 0; green, 0; blue, 0 }  ,fill opacity=1 ][line width=1.5]  (316.5,219.5) .. controls (316.5,214.53) and (320.53,210.5) .. (325.5,210.5) .. controls (330.47,210.5) and (334.5,214.53) .. (334.5,219.5) .. controls (334.5,224.47) and (330.47,228.5) .. (325.5,228.5) .. controls (320.53,228.5) and (316.5,224.47) .. (316.5,219.5) -- cycle ;
\draw  [fill={rgb, 255:red, 0; green, 0; blue, 0 }  ,fill opacity=1 ][line width=1.5]  (400.25,219.5) .. controls (400.25,214.53) and (404.28,210.5) .. (409.25,210.5) .. controls (414.22,210.5) and (418.25,214.53) .. (418.25,219.5) .. controls (418.25,224.47) and (414.22,228.5) .. (409.25,228.5) .. controls (404.28,228.5) and (400.25,224.47) .. (400.25,219.5) -- cycle ;
\draw  [fill={rgb, 255:red, 0; green, 0; blue, 0 }  ,fill opacity=1 ][line width=1.5]  (484,219.5) .. controls (484,214.53) and (488.03,210.5) .. (493,210.5) .. controls (497.97,210.5) and (502,214.53) .. (502,219.5) .. controls (502,224.47) and (497.97,228.5) .. (493,228.5) .. controls (488.03,228.5) and (484,224.47) .. (484,219.5) -- cycle ;
\draw [line width=1.5]    (158,210.5) .. controls (177.5,181.74) and (214.35,174.84) .. (236.34,203.71) ;
\draw [shift={(238,206)}, rotate = 235.38] [color={rgb, 255:red, 0; green, 0; blue, 0 }  ][line width=1.5]    (14.21,-6.37) .. controls (9.04,-2.99) and (4.3,-0.87) .. (0,0) .. controls (4.3,0.87) and (9.04,2.99) .. (14.21,6.37)   ;
\draw [line width=1.5]    (241.75,210.5) .. controls (260,156) and (280,199) .. (280,217) .. controls (280,234.73) and (265.69,286.9) .. (246.86,234.47) ;
\draw [shift={(246,232)}, rotate = 71.19] [color={rgb, 255:red, 0; green, 0; blue, 0 }  ][line width=1.5]    (14.21,-6.37) .. controls (9.04,-2.99) and (4.3,-0.87) .. (0,0) .. controls (4.3,0.87) and (9.04,2.99) .. (14.21,6.37)   ;
\draw [line width=1.5]    (325.5,210.5) .. controls (345,181.74) and (384.23,173.89) .. (406.33,202.71) ;
\draw [shift={(408,205)}, rotate = 235.38] [color={rgb, 255:red, 0; green, 0; blue, 0 }  ][line width=1.5]    (14.21,-6.37) .. controls (9.04,-2.99) and (4.3,-0.87) .. (0,0) .. controls (4.3,0.87) and (9.04,2.99) .. (14.21,6.37)   ;
\draw [line width=1.5]    (409.25,210.5) .. controls (428.75,181.74) and (466.31,174.84) .. (488.34,203.71) ;
\draw [shift={(490,206)}, rotate = 235.38] [color={rgb, 255:red, 0; green, 0; blue, 0 }  ][line width=1.5]    (14.21,-6.37) .. controls (9.04,-2.99) and (4.3,-0.87) .. (0,0) .. controls (4.3,0.87) and (9.04,2.99) .. (14.21,6.37)   ;
\draw [line width=1.5]    (162.77,235.49) .. controls (182.38,261.56) and (220.65,261.96) .. (241.75,228.5) ;
\draw [shift={(161,233)}, rotate = 56.31] [color={rgb, 255:red, 0; green, 0; blue, 0 }  ][line width=1.5]    (14.21,-6.37) .. controls (9.04,-2.99) and (4.3,-0.87) .. (0,0) .. controls (4.3,0.87) and (9.04,2.99) .. (14.21,6.37)   ;
\draw [line width=1.5]    (324.3,201.62) .. controls (312.65,148.67) and (292,201.34) .. (292,218) .. controls (292,235) and (311,284) .. (325.5,228.5) ;
\draw [shift={(325,205)}, rotate = 258.88] [color={rgb, 255:red, 0; green, 0; blue, 0 }  ][line width=1.5]    (14.21,-6.37) .. controls (9.04,-2.99) and (4.3,-0.87) .. (0,0) .. controls (4.3,0.87) and (9.04,2.99) .. (14.21,6.37)   ;
\draw [line width=1.5]    (330.76,235.49) .. controls (350.35,261.56) and (388.15,261.96) .. (409.25,228.5) ;
\draw [shift={(329,233)}, rotate = 56.31] [color={rgb, 255:red, 0; green, 0; blue, 0 }  ][line width=1.5]    (14.21,-6.37) .. controls (9.04,-2.99) and (4.3,-0.87) .. (0,0) .. controls (4.3,0.87) and (9.04,2.99) .. (14.21,6.37)   ;
\draw [line width=1.5]    (414.76,235.49) .. controls (434.34,261.56) and (471.9,261.96) .. (493,228.5) ;
\draw [shift={(413,233)}, rotate = 56.31] [color={rgb, 255:red, 0; green, 0; blue, 0 }  ][line width=1.5]    (14.21,-6.37) .. controls (9.04,-2.99) and (4.3,-0.87) .. (0,0) .. controls (4.3,0.87) and (9.04,2.99) .. (14.21,6.37)   ;

\draw (131,37) node [anchor=north west][inner sep=0.75pt]  [font=\large] [align=left] {(a)};
\draw (336,79) node [anchor=north west][inner sep=0.75pt]   [align=left] {$\displaystyle w$};
\draw (254,80) node [anchor=north west][inner sep=0.75pt]   [align=left] {$\displaystyle v$};
\draw (131,172) node [anchor=north west][inner sep=0.75pt]  [font=\large] [align=left] {(b)};
\draw (254,214) node [anchor=north west][inner sep=0.75pt]   [align=left] {$\displaystyle v$};
\draw (339,211) node [anchor=north west][inner sep=0.75pt]   [align=left] {$\displaystyle w$};

\end{tikzpicture}